\documentclass[a4paper,11pt]{article}
\usepackage{pos}

\newcommand{\alphas}{\alpha_s}
\newcommand{\as}{\alpha_s}
\newcommand{\rd}{\text{d}}
\newcommand{\order}[1]{\mathcal{O}(#1)}

\newcommand{\calF}{\mathcal{F}}
\newcommand{\expo}[1]{\ensuremath{\mathrm{e}^{#1}}}

\usepackage{comment}
\title{Precise Predictions for Hadronic Higgs Decays}

\author*[a]{Elliot Fox}

\affiliation[a]{Institute for Particle Physics Phenomenology, Department of Physics, University of Durham, Durham, DH1 3LE, UK}

\emailAdd{f.author@inst.edu}
\emailAdd{s.author@univ.country}

\abstract{The prospect of future electron-positron colliders operating as "Higgs factories" in a clean experimental environment presents one of the most promising avenues for Higgs precision measurements. In order to capitalise on this, we need to have good theoretical control over these observables. In this talk, I will report on recent calculations in Hadronic Higgs decays, focusing in particular on the variations between the dominant $H\to b \bar{b}$ channel via a Yukawa interaction, and the sub-dominant $H\to gg$ channel. Using the newly-developed "generalised antenna formalism", we have been able to calculate jet-rates and classical QCD event-shape observables up to NNLO accuracy, providing us with the means to quantify the differences between the two decay modes. For a subset of observables, we also match these NNLO results to NNLL resummation to obtain valid predictions even in the back-to-back limit.
	
}

\FullConference{17th International Symposium on Radiative Corrections: Applications of Quantum Field Theory to Phenomenology (RADCOR2025)\\
5-10 October 2025\\
Puri, India\\}

\begin{document}
\maketitle
\section{Introduction}
At the Large Hadron Collider (LHC), the Higgs boson has been discovered by both the ATLAS ~\cite{ATLAS:2012yve} and CMS ~\cite{CMS:2012qbp} experiments, and its dominant decay channel to $b\bar{b}$ via a Yukawa interaction 
has been observed in the relatively clean VH production channel. Measurements of the sub-dominant decay to two gluons via a heavy quark loop 
however have not yet been possible due to the large irreducible QCD background present at hadron-hadron colliders. For future electron-positron colliders such as FCC-ee~\cite{Abada:2019zxq} and CEPC~\cite{CEPCStudyGroup:2018ghi} the outlook is more promising. They will operate in a "Higgs-factory" mode, meaning the centre-of-mass energy will be enough to produce a large number of Higgs bosons, comparable to the number of Z bosons which were produced at LEP~\cite{ALEPH:2003obs}. This, coupled with the much cleaner lepton-lepton collider environment free from initial-state QCD radiation, means the experimental uncertainties on Higgs observables are expected to be cut down to the per mille level, and a precise theory description is needed in order to capitalise on this. 

In this work, we report on the calculation of precise predictions for Higgs decay observables, including jet rates and three-jet event shapes, primarily focussing on the Thrust distribution.  
\section{Hadronic Higgs Decays}
\label{Sec:HadronicHiggsDecays}
The dominant channel for the hadronic decay of the Higgs is to a quark-antiquark pair via a Yukawa interaction, and corresponds to $\sim 85\%$ of all hadronic Higgs decay events. The sub-dominant decay channel is to gluons via a heavy quark loop and, contributes $\sim 15\%$. We consider this decay mode in an effective field theory in which heavy quarks are treated as infinitely massive and are integrated out. The N$^{k}$LO inclusive width for the decay mode X=$q\bar{q},gg$ is given by
\begin{align}
	\Gamma^{(k)}_{H\to X} &= \Gamma^{(0)}_{H\to X}\, \left(1+\sum\limits_{n=1}^k\alphas^n C_{X}^{(n)}\right)\,,
	\label{eq:ratesNkLO}
\end{align}
where  $\Gamma^{(0)}_{H\to X}$ is the LO width, and is given along with the $C_X^{(n)}$ up to n=4 in  \cite{Herzog:2017dtz}.
We note that although we consider quarks with non-zero mass in their respective Yukawa couplings (in practice the only non-negligible contributions are from bottom and charm quarks), light quarks are treated as massless in the rest of the calculation. This assumption means that to all order in perturbation theory the two modes do not interfere, and hence splitting into the two independent channels is meaningful. When masses are included we verified that the Born-level interference contributes well below the percent level for all observables we consider.

Throughout this work we consider the decay on an on-shell Higgs boson with mass $m_H=125.09$ GeV, and work in the $G_{\mu}$ scheme with constant electroweak parameters 
\begin{equation}
	G_F=1.1664\cdot 10^{-5}\text{GeV}^{-2},\quad m_Z = 91.200 \text{GeV}
\end{equation}
 giving a Higgs vacuum expectation of $v=246.22$GeV. We take $\alphas(m_z)=0.11800$, and consider two-loop running yielding $\alphas(m_H)=0.11263$. Top, bottom and charm quark masses are taken in the $\bar{MS}$ scheme, and are given by $m_t(m_H)=166.48$ GeV, $y_b(m_H)=m_b(m_H)/v=0.013309$ and $y_c(m_H)=m_c(m_H)/v=0.0024629$.   
\section{Calculation Setup}
The calculation requires matrix elements for the decay of a Higgs boson to up to five partons at tree level~\cite{DelDuca:2004wt,Anastasiou:2011qx,DelDuca:2015zqa,Mondini:2019vub,Mondini:2019gid}, up to four partons at one-loop ~\cite{Dixon:2009uk,Badger:2009hw,Badger:2009vh,Anastasiou:2011qx,DelDuca:2015zqa,Mondini:2019vub,Mondini:2019gid} and three partons at two-loop. For the latter case, the $H\to gg$ two-loop matrix elements are (up to crossing) the same as those computed in~\cite{Gehrmann:2011aa}  and used in the NNLO calculation for Higgs plus jet production at hadron colliders  ~\cite{Chen:2014gva,Chen:2016zka}. In the $H\to q\bar{q}$ case, the two-loop matrix elements were originally calculated in ~\cite{Ahmed:2014pka,Mondini:2019vub}, and we recomputed them, achieving full numerical agreement with ~\cite{Mondini:2019vub}.

To treat the infrared singularities which appear at intermediate stages of higher-order calculations we use the antenna subtraction method 
~\cite{Gehrmann-DeRidder:2005btv}, which uses \textit{antenna functions} to construct subtraction terms that remove the divergent behaviour of real matrix elements in unresolved limits. After analytic integration over the antenna phase space, they are then added back at more virtual layers and remove the explicit poles which exist in loop amplitudes. In particular, we rely on \textit{idealised antenna functions} constructed from a target set of infrared limits by applying the algorithm detailed in \cite{Braun-White:2023sgd,Braun-White:2023zwd}, and the extension to \textit{generalised antenna functions} outlined in \cite{Fox:2024bfp}. These recent advancements significantly simplifies the construction of subtraction terms, and the resulting implementation is much more efficient with respect to the traditional method.

The IR-finite remainders are numerically integrated with the Monte Carlo event generator \textsc{NNLOjet} \cite{NNLOJET:2025rno}, suitably adapted for the implementation of generalised antenna functions. 
We validated our LO and NLO results for the production of three and four jets against \textsc{Eerad3}~\cite{Gehrmann-DeRidder:2014hxk,Coloretti:2022jcl,Gehrmann-DeRidder:2023uld} and find very good agreement. For the Yukawa-induced mode we reproduce the results of~\cite{Mondini:2019vub,Mondini:2019gid}.
\label{sec:CalcSetup}
\section{Jet Rates}
\label{Sec:JetRates}
The first observable we consider is the n-jet rate (for n=2,3,4,5), which is defined as the fraction of hadronic Higgs decay events that are clustered by a given jet algorithm into n-jets. Specifically, we consider the Durham jet algorithm, and our results are differential in the jet resolution parameter $y_{cut}$ which we define below. The discussion and results follow what is reported in \cite{Fox:2025cuz}.

The Durham jet algorithm takes in $n_{part}$ particles and clusters them into $n_{jets}\leq n_{part}$ jets as follows. For each pair of particles (i,j), the distance measure 
	\begin{eqnarray}
		y_{ij}=\dfrac{2\text{min}(E_i^2,E_j^2)(1-\cos(\theta_{ij}))}{Q^2}
	\end{eqnarray}
	is calculated, where Q=$\sum_kE_k$ the sum of the energies of all final state particles, and $\theta_{ij}$ is the angle between the particles i and j. The particle pair (i,j) with the smallest distance measure is identified, and if $y_{ij}<y_{cut}$, particles i and j are clustered together by adding together their four-momenta. The algorithm then repeats until all $y_{ij}>y_{cut}$, at which point the final state is classified as an n-jet event.  Clearly, the smaller the value of $y_{cut}$, the earlier the algorithm will terminate, and hence the more distinct jets are resolved, and vice versa.

We define the n-jet rate at $\mathcal{O}(\alphas^k)$ differential in $y_{cut}$ as 
\begin{eqnarray}
	R_X^{(k)}(n,y_{cut})=\dfrac{\Gamma_{H\to X}^{(k)}(n,y_{cut})}{\Gamma_{H\to X}^{(k)}},\quad \text{with }X=gg,q\bar{q}.
\end{eqnarray}
The three-, four- and five-jet rates at NNLO, NLO and LO respectively we calculate directly, and the two-jet rate at N$^3$LO we infer from unitarity, i.e. that at any order the rates sum to unity.
\begin{figure*}[t]
	\centering
	\label{fig:3jetNNLOa}\includegraphics[width=0.26\textwidth]{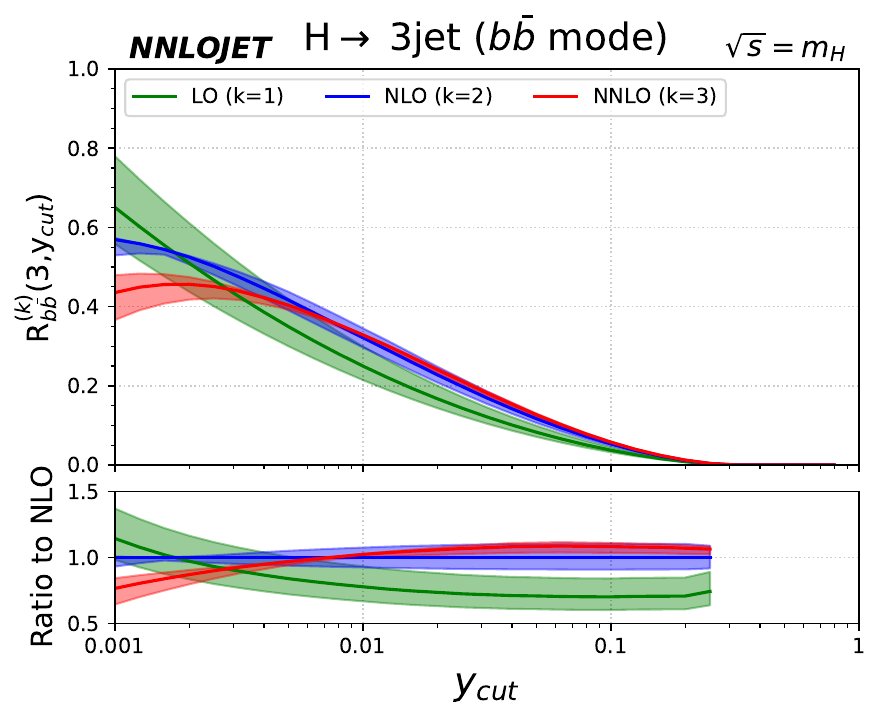}
	\label{fig:3jetNNLOb}\includegraphics[width=0.26\textwidth]{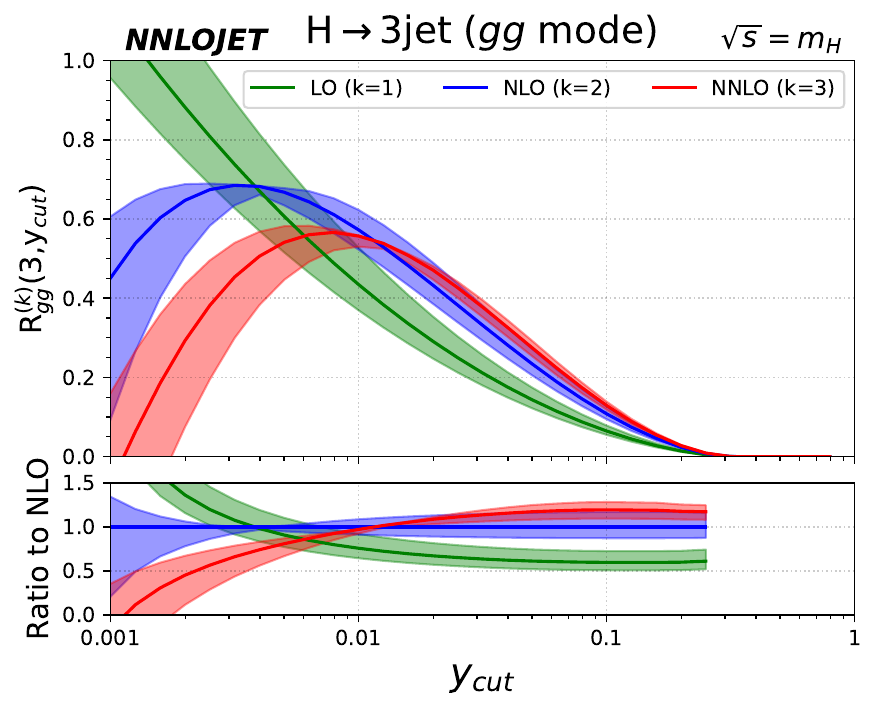}
	\label{fig:3jetNNLOc}\includegraphics[width=0.26\textwidth]{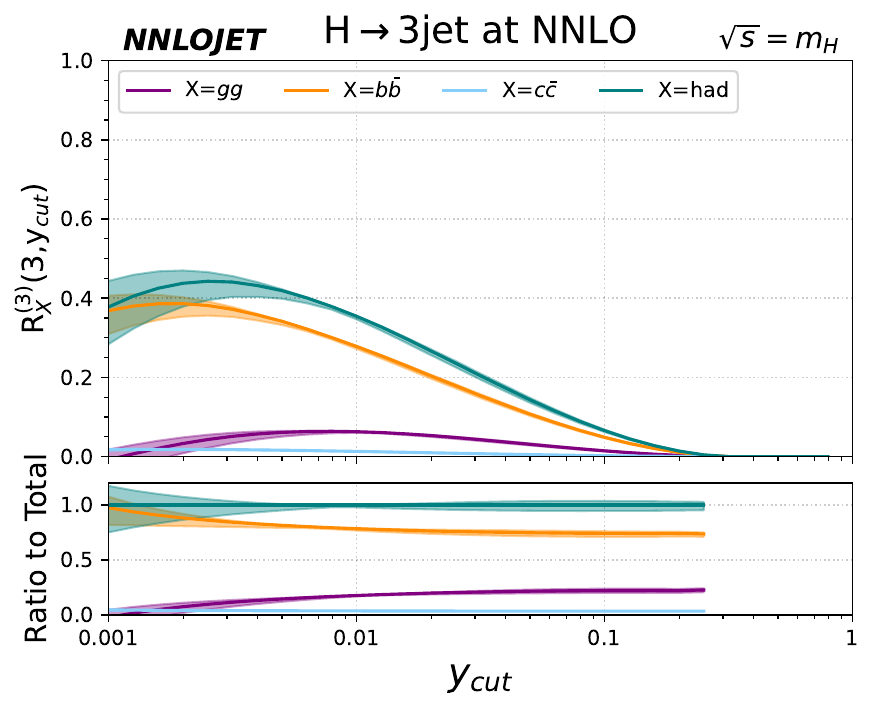}
	\caption{Normalised three-jet decay rate at LO (green), NLO (blue) and NNLO (red) for Higgs decay to bottom quarks (left), Higgs to gluons (centre), and the weighted sum of the different decay modes (right). The corresponding inclusive decay rate is used as normalisation (see text).}
	\label{fig:3jetNNLO}
\end{figure*} 
In Fig. \ref{fig:3jetNNLO} we present the 3-jet rate at LO, NLO and NNLO for Higgs decay to quarks (Frame 1), Higgs decay to gluons (Frame 2) and the total Hadronic Higgs decay (Frame 3). We obtain this latter plot as a sum of the three-jet rate for Higgs decays to bottom quarks, charm quarks and gluons, normalised to the total hadronic width at N$^k$LO given by
\begin{eqnarray}
	\Gamma^{(k)}_{H\to \text{had}}=\Gamma^{(k)}_{H\to b\bar{b}}+\Gamma^{(k)}_{H\to c\bar{c}}+\Gamma^{(k)}_{H\to gg}.
	\label{eq:gamma_tot}
\end{eqnarray}
To estimate theory uncertainties, we vary the renormalisation scale $\mu_R$ between $[m_h/2,2m_H]$ in a correlated manner between numerator and denominator. We observe that in general, both $H\to q\bar{q}$ and $H\to gg$ exhibit good perturbative convergence for large values of $y_{cut}$,  demonstrated by the scale variation bands overlapping as we go to higher orders. As we go to lower values of $y_{cut}$, this perturbative convergence breaks down due to the presence of large-logarithmic contributions, the onset of which occurs at higher values of $y_{cut}$ in the $H\to gg$ channel than the $H\to q\bar{q}$ channel. In fact, for $y_{cut}<0.001$ the $H\to gg$ prediction is negative, which is clearly unphysical. Similarly, the peak for the $H\to gg$ channel occurs at a larger value of $y_{cut}$ ($\sim 0.007$) than for the $H\to q\bar{q}$ channel ($\sim 0.002$). In Frame 3 we observe that for large values of $y_{cut}$ the sensitivity to the $H\to gg$ channel is enhanced (e.g. $\sim25\%$ for $y_{cut}= 0.1$), whereas at $y_{cut}=0.001$ the total three-jet rate is $\sim 40\%$, but the contribution from the $H\to gg$ channel vanishes.
\section{The Thrust Distribution at NNLO}
\label{Sec:ThrustNNLO}
\begin{figure}[htbp]
	\centering
	\includegraphics[width=0.26\linewidth]{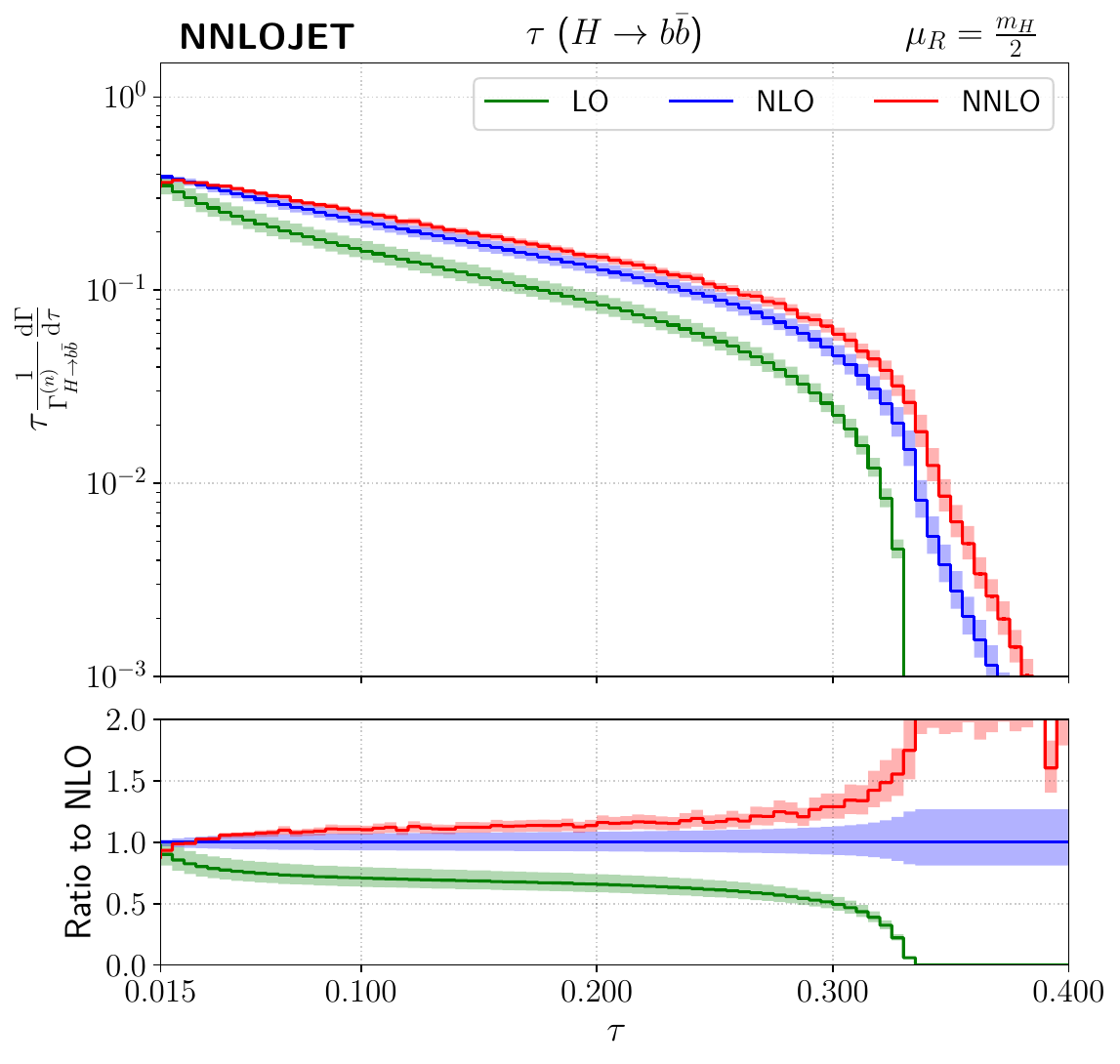}
	\includegraphics[width=0.26\linewidth]{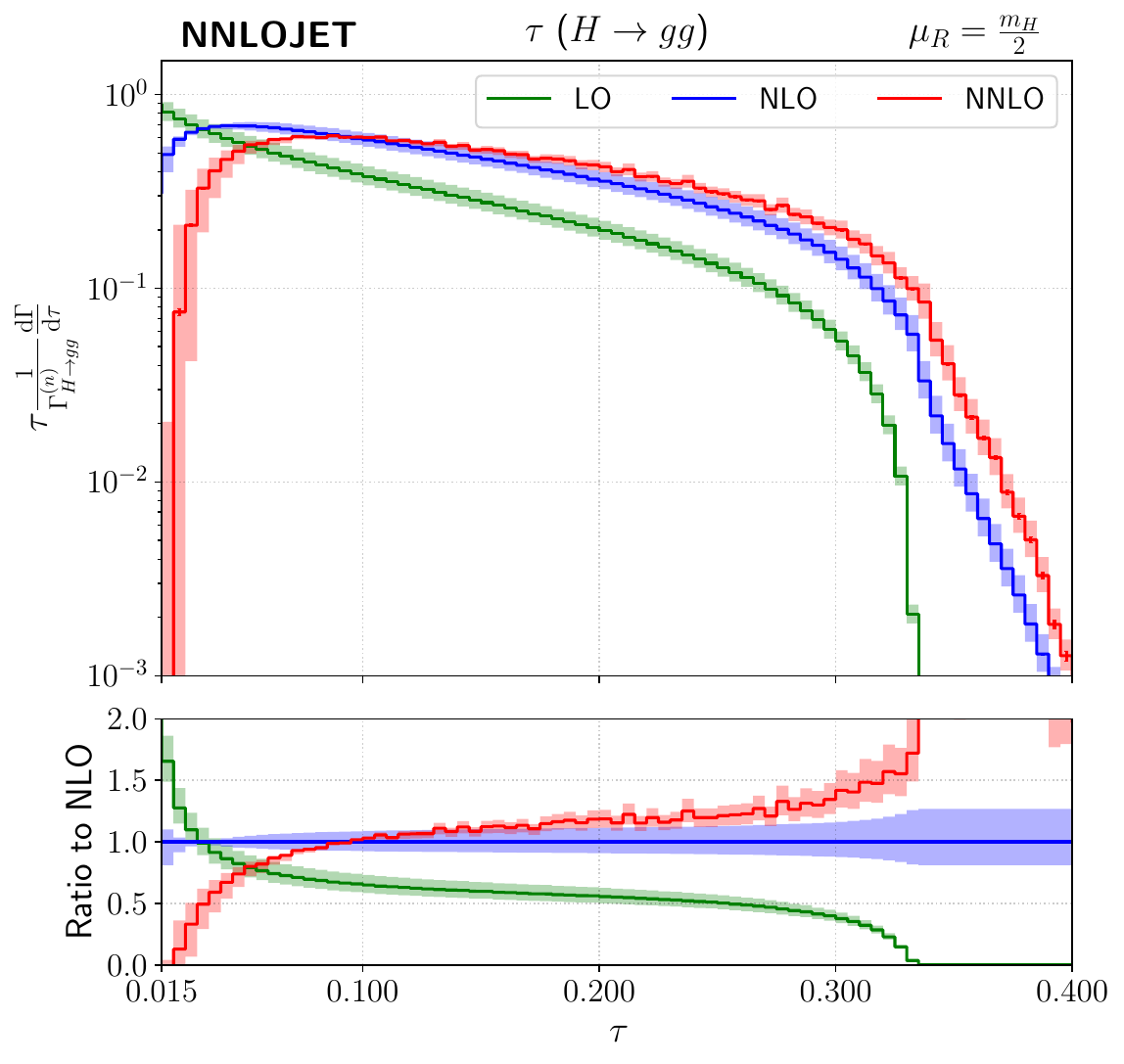}
	\includegraphics[width=0.26\linewidth]{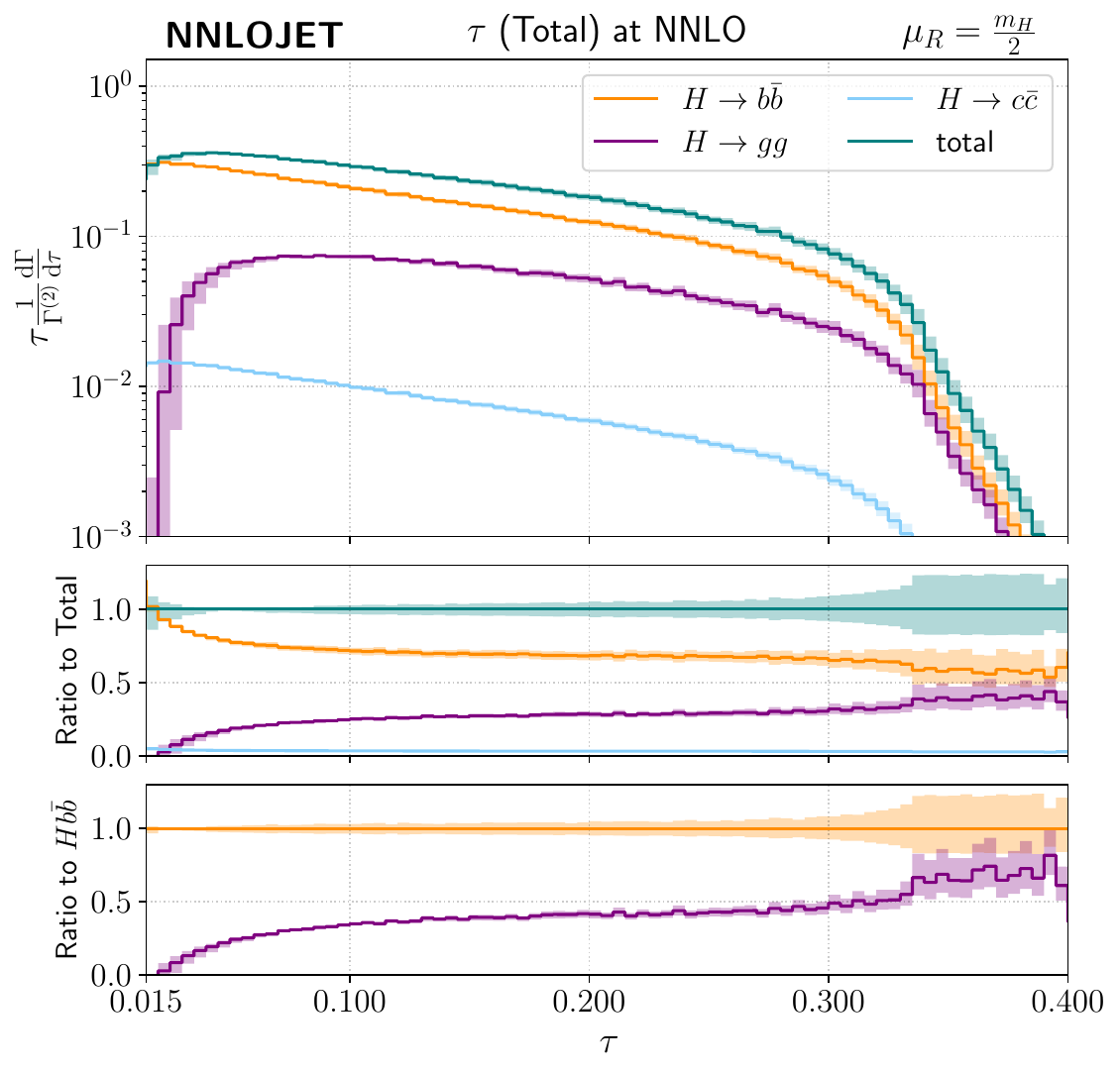}
	\caption{Thrust distributions for the $H\to b\bar{b}$ channel (top row), the $H\to gg$ channel (middle row) and the sum of all decay modes at NNLO (bottom row). In Frames 1 and 2, the LO (green), NLO (blue), and NNLO (red) are shown. In Frame 3, the individual contributions of the $H\to b\bar{b}$ (orange), $H\to gg$ (purple), and the $H\to c\bar{c}$ (light blue) channels are displayed alongside the sum (teal).}
	\label{fig:tau_HJM}
\end{figure}

The implementation in \textsc{NNLOjet} used to calculate the jet-rate observables as described above can also be applied to predictions for classical event shape observables, the results of which are presented in \cite{Fox:2025qmp}. Here we focus specifically on the thrust observable, defined as $\tau=1-T$, where
\begin{eqnarray}
	T = \max\limits_{\vec{n}}\left(\dfrac{\sum\limits_i |\vec{p}_i\cdot\vec{n}|}{\sum\limits_i |\vec{p}_i|} \right) \,.
\end{eqnarray}
We present results for the differential decay rate normalised to the NNLO inclusive decay rate and expanded systematically in $\alphas$, or concretely
\begin{eqnarray}\label{eq:exp2}
	\frac{1}{\Gamma^{(2)}_{H\to X}(m_H,\mu_R)}\frac{\rd \Gamma_{H\to X}(m_H,\mu_R)}{\rd \tau}&=&\nonumber \\ &&\hspace{-6cm}\left(\frac{\alphas(\mu_R)}{2\uppi}\right)\frac{\rd \bar{A}_X(\mu_R)}{\rd \tau} + \left(\frac{\alphas(\mu_R)}{2\uppi}\right)^2\frac{\rd \bar{B}_X(\mu_R)}{\rd \tau} + \left(\frac{\alphas(\mu_R)}{2\uppi}\right)^3\frac{\rd \bar{C}_X(\mu_R)}{\rd \tau} + \mathcal{O}(\alpha_s^4) \,.
	\label{eq:rateDiffNorm}
\end{eqnarray}
The coefficients $\bar{A}$, $\bar{B}$, and $\bar{C}$ are given in terms of Eq.~\eqref{eq:ratesNkLO} by
\begin{equation}
	\label{eq:Abar}
	\bar{A} _X= A_X\,, \quad \bar{B}_X = B_X - C_X^{(1)}A_X\,, \quad \bar{C}_X = C_X - C^{(1)}_XB_X + \left(\left(C_X^{(1)}\right)^2-C_X^{(2)}\right)A_X \,,
\end{equation} 
where A, B and C are the LO, NLO and NNLO coefficients normalised to the LO inclusive width.
In Fig.\ref{fig:tau_HJM}, we present the results for $H\to q\bar{q},gg$ and the sum over all channels with $\mu_R$ varied between $m_H/4$ and $mH$. In both Frames 1 and 2 we see that at $\tau=1/3$ there is a Sudakov shoulder where the LO distribution is identically zero, and for values of $\tau>1/3$ the NNLO correction is enhanced. In Frame 3, we see that in the region $\tau\in[0.1,0.3]$ the fraction of the summed distribution from the Yukawa/gluonic channel remains roughly constant at $70\%$ and $30\%$ respectively. For values of $\tau$ larger than this the gluonic mode is enhanced up to more than a third of the total, whereas for smaller $\tau$ values its contribution reduces.  In general we observe that even in the bulk region the perturbative convergence is poor, as the scale variation bands do not overlap. This will be remedied by the inclusion of resummation effects discussed in the next section.  
\section{The Thrust Distribution at NNLL matched to NNLO}
\label{Sec:ThrustNNLL}
In general for a given three-jet event shape y, in the limit $y\to 0$ any fixed order prediction will generate large logarithmic terms of the form $\alphas^nL^m$ where L=-log(y). Since this is no longer a small expansion parameter, perturbation theory breaks down and leads to unphysical predictions, for example in Fig.\ref{fig:tau_HJM} where for low values of $\tau$ the differential cross-section in the gluonic mode becomes negative. In order to restore predictive power, the large logarithmic contributions must be resummed to all orders. In this section we do this for the $y=\tau$ observable, and then match the results to our fixed order predictions in to obtain distributions valid across the full range of $\tau$ values, following the discussion given in \cite{Fox:2025txz}.

\begin{figure}[t]
	\centering
	\includegraphics[width=0.26\textwidth]{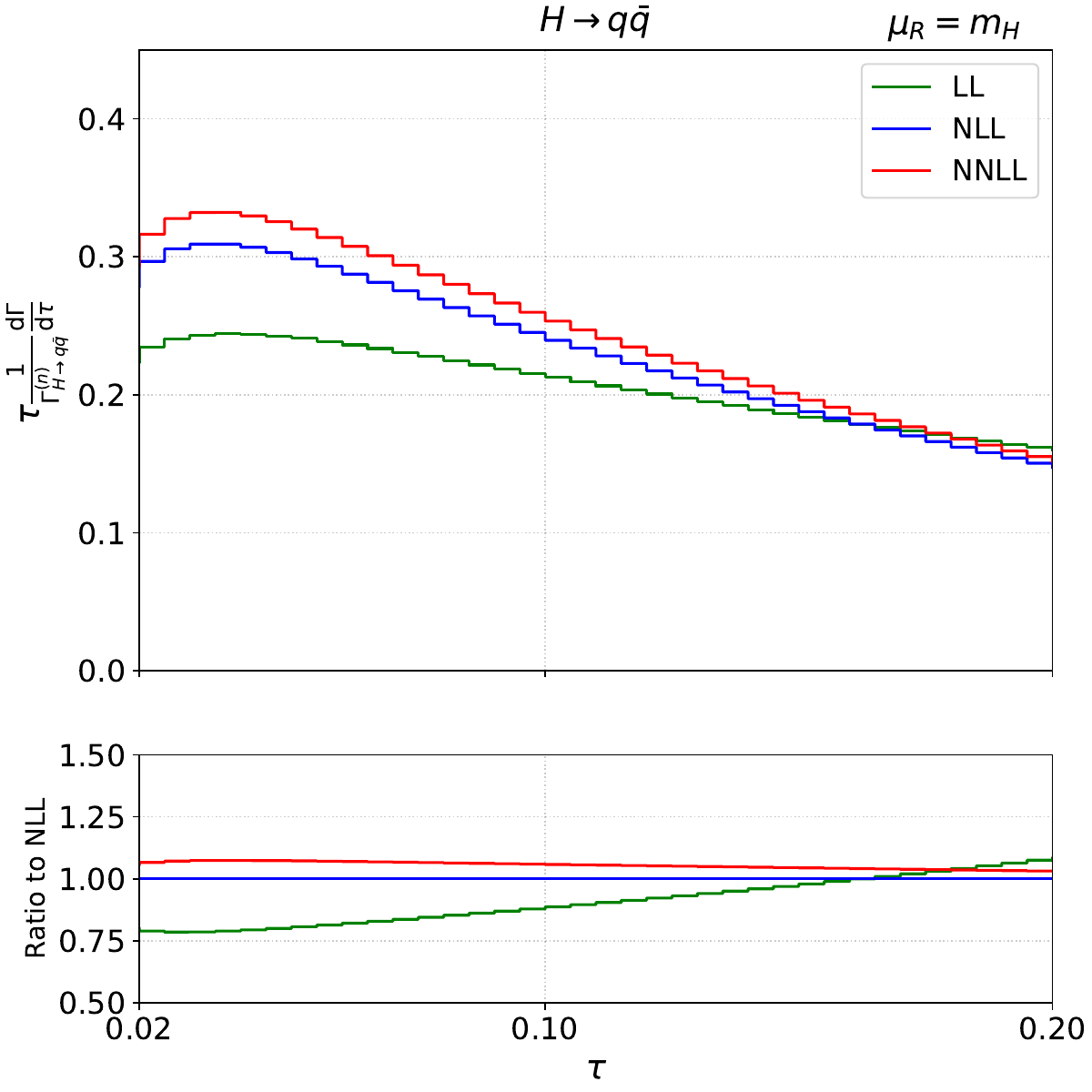}
	\includegraphics[width=0.26\textwidth]{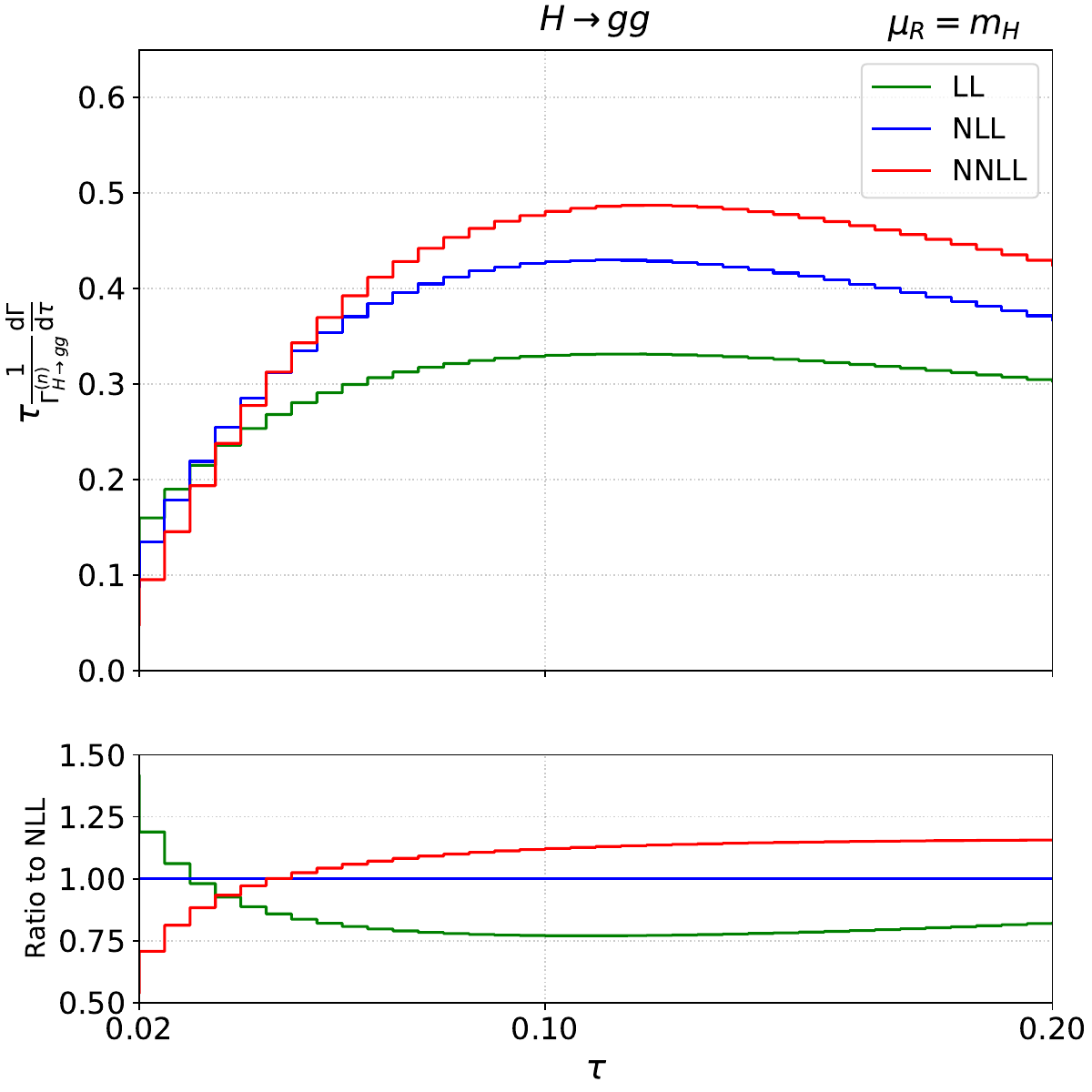}
	\caption{Comparison of resummed results for thrust at LL (green), NLL (blue), and NNLL (red) in $H\to q\bar{q}$ decays (left) and $H\to gg$ decays (right). }
	\label{fig:resummed}
\end{figure}

\begin{figure}[t]
	\centering
	\includegraphics[width=0.26\textwidth]{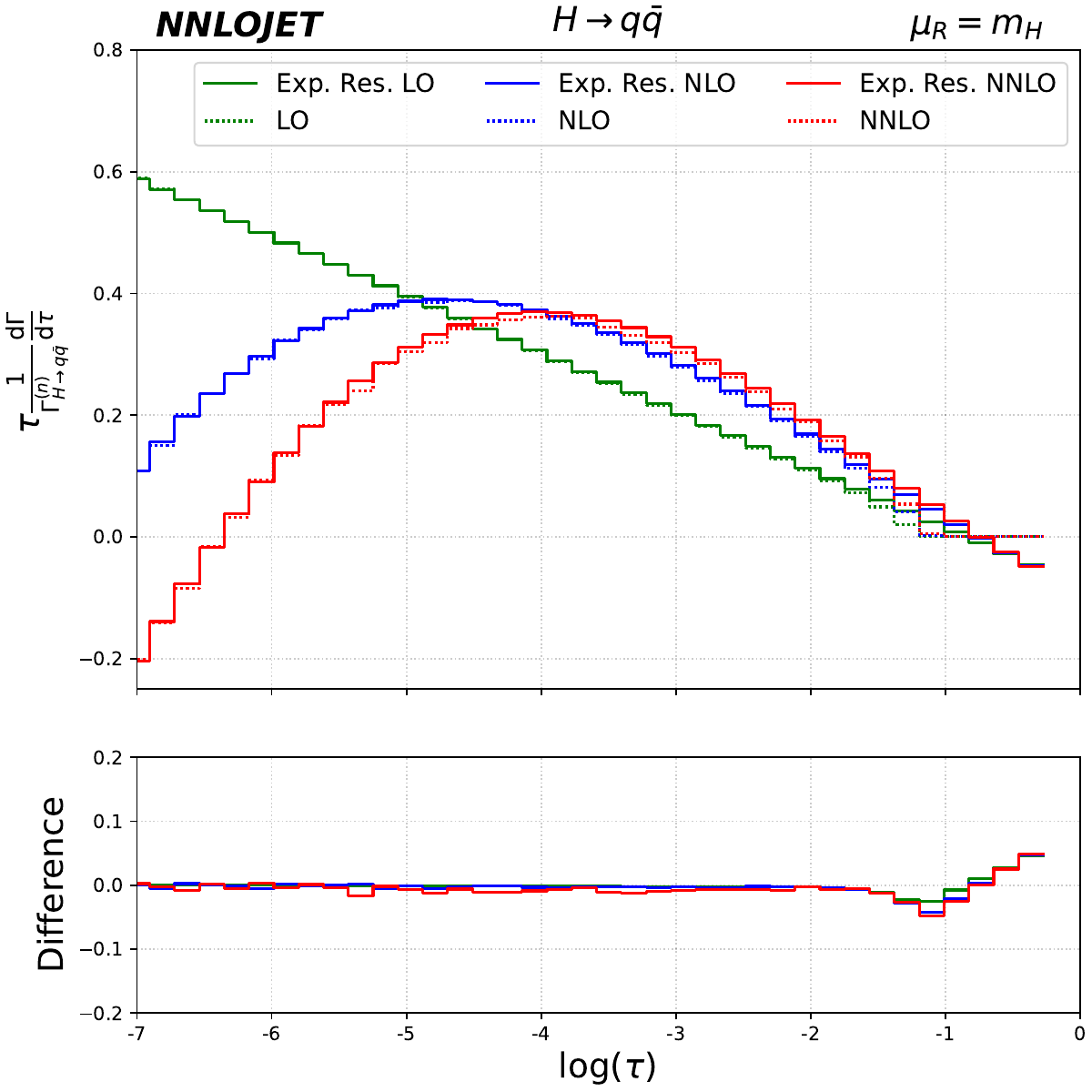}
	\includegraphics[width=0.26\textwidth]{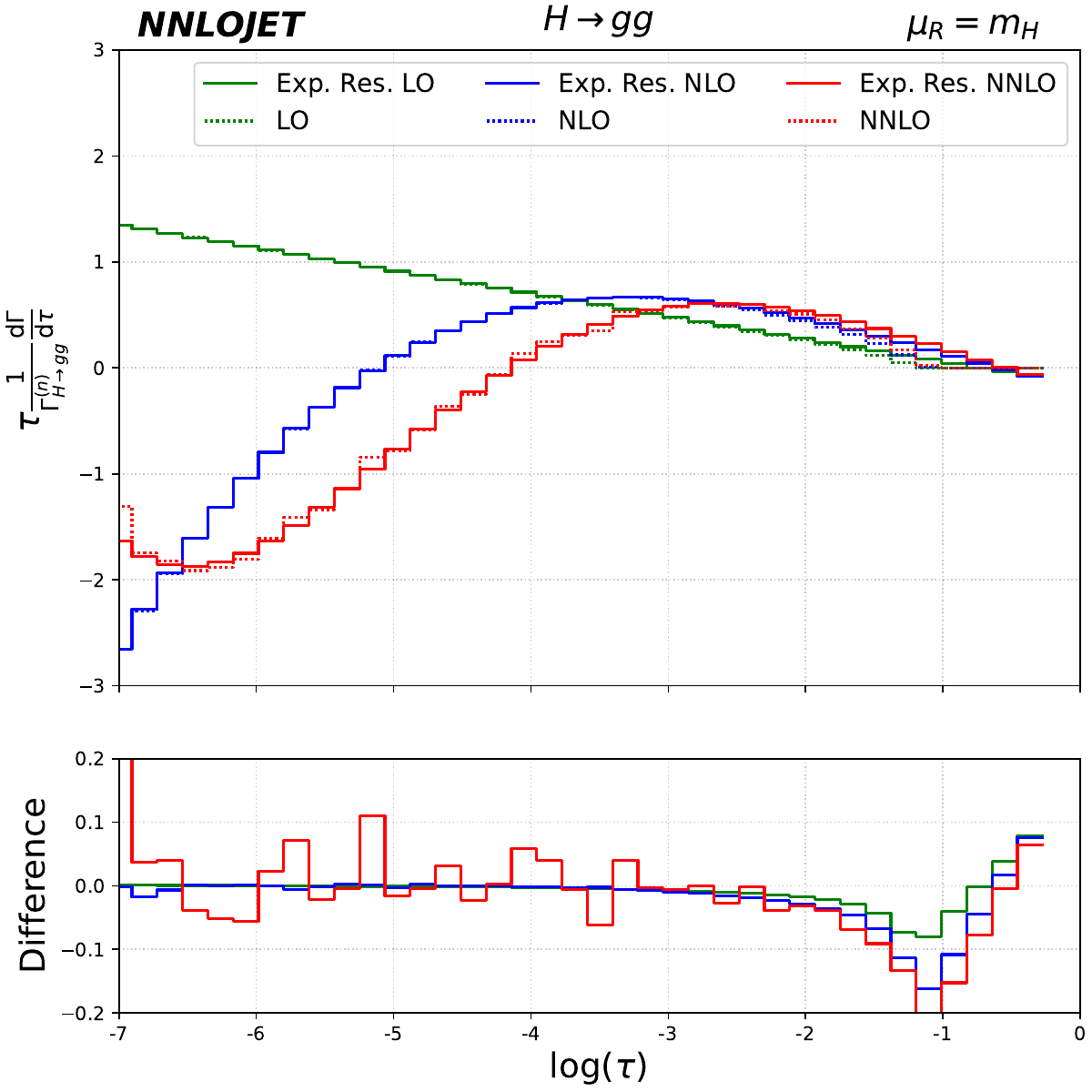}
	\caption{Comparison between the expansion of the resummation formula (solid lines) and the fixed-order results (dashed lines) up to $\mathcal{O}(\alpha_s)$ (LO, green), $\mathcal{O}(\alpha_s^2)$ (NLO, blue) and $\mathcal{O}(\alpha_s^3)$ (NNLO, red). The difference between  the fixed-order and the expansion of the resummation formula is shown in the lower frames.}
	\label{fig:check}
\end{figure}

To formulate the resummation of the $\tau$ event-shape variable we work at the level of the cumulant of the differential decay rate distribution, defined as
\begin{align}
	\Sigma_{H\to X}(\tau) &= \frac{1}{\Gamma_{H\to X}}\int\limits^{\tau}_0\, \frac{\rd \Gamma_{H\to X}}{\rd \tau'}\, \rd \tau  \,.
	\label{eq:cumulant1}
\end{align}
This admits a fixed order expansion
\begin{equation}
	\Sigma^{NNLO}_{H\to X}(\tau) = 1 + \as\mathcal{A}(\tau) + \as^2\mathcal{B}(\tau) + \as^3\mathcal{C}(\tau) + \order{\as^4}\,,
	\label{eq:cumulantFO}
\end{equation}
where $\mathcal{A},\mathcal{B},\mathcal{C}$ can be computed from $\bar{A},\bar{B},\bar{C}$ as defined in Eq.\ref{eq:Abar}, and provides reliable results for intermediate to large values of $\tau$. Crucially, it also admits a description valid in the $\tau\to0$ limit where all the large logarithmic terms have been resummed to all orders in $\alphas$ given by
\begin{align}
	\Sigma_{H\to X}(y) = \left(1+\sum\limits_n \as^n c_{n}^{(H\to X)}\right)\expo{-R_X(\alphas L)}\calF_X(R_X^\prime(\alphas L))\,,
	\label{eq:cumulant}
\end{align}
where $R_X$ is the Sudakov radiator describing single-emission contributions and the $\calF_X$ function accounts for multiple emissions. Together they encode the all-order behaviour of $\Sigma(y)$, and depend only on the observable y and the particle type of the hard radiator. The coefficients $c_n^(H\to X)$ meanwhile induce an overall normalisation, and are both non-logarithmic and process dependent.
Just as in the fixed order case, this resummed distribution is systematically improvable, as we can resum the leading logarithms (LL) which contribute in the exponent as $\alphas^nL^{n+1}$, next-to-leading logarithms (NLL) which contribute as $\alphas^nL^{n}$, next-to-next-to leading logarithms (NNLL) which contribute as $\alphas^nL^{n-1}$, and in principle beyond. We compute the necessary ingredients in Eq.\ref{eq:cumulant} using the ARES formalism~\cite{Banfi:2014sua,Banfi:2018mcq} up to NNLL, and give the results in detail in \cite{Fox:2025txz}. The differential decay rate can then be inferred from the cumulant as 
	\begin{eqnarray}                                                                                                                                                                                                                                                                                                              
		\dfrac{\text{d}\Sigma_{H\to X}(\tau)}{\text{d}\log(\tau)}=\tau\dfrac{1}{\Gamma^{(k)}_{H\to X}(m_H,\mu_R)}\dfrac{\text{d}\Gamma_{H\to X}(m_H,\mu_R)}{\text{d}\tau}.                                                                                                                                                                    
	\end{eqnarray} 
In Fig.\ref{fig:resummed} we show the resummed distribution and LL, NLL and NNLL. It is clear that the unphysical features of the fixed-order distribution for low values of y (e.g. predictions becoming negative) are not present in the resummed distribution. 
In order to obtain results valid for all values of $\tau$, we would like to combine the resummed results in the low $\tau$ region and the fixed-order results in the high $\tau$ region: this comes under the name of matching. To do this, we first consider that the resummed cumulant defined in Eq.\ref{eq:cumulant} can be expanded and truncated to a given order in $\alphas$ as 
\begin{align}
	\Sigma_\text{LL}(y) &= \exp\left\{G_{12}a_sL^2 + G_{23}a_s^2L^3 + G_{34}a_s^3L^4 + \ldots\right\}\,, \\
	\Sigma_\text{NLL}(y) &= \exp\left\{G_{11}a_sL + G_{22}a_s^2L^2 + G_{33}a_s^3L^3 + \ldots\right\} \,, \\
	\Sigma_\text{NNLL}(y) &= \exp\left\{G_{21}a_s^2L + G_{32}a_s^3L^2 + \ldots\right\} \,.
\end{align}
In Fig.\ref{fig:check} we compare this expansion with the fixed order result, and find that the divergence as $y\to 0$ predicted by truncated the resummed distribution matches very well with the actual divergence exhibited by the fixed order distribution. This serves as a strong check of both our resummed and fixed order results, as well as providing us with the means to fit the two descriptions together.

Specifically, we employ the log-R matching scheme ~\cite{Catani:1992ua,Jones:2003yv}, which is defined as
\begin{equation}
	\label{eq:matching}
	\begin{split}
		\log\left(\Sigma_{NNLO+NNLL}(\tau)\right) &= -R_\text{NNLL}(\tau) + \log\left(\calF_\text{NNLL}(\tau)\right) \\
		&\hspace{-2.5cm}+a_s\left(\mathcal{A}(\tau) - G_{11}L - G_{12}L^2\right)+a_s^2\left(\mathcal{B}(\tau) - \frac{1}{2}\mathcal{A}(y)^2 - G_{21}L - G_{22}L^2 - G_{23}L^3\right) \\
		&\hspace{-2.5cm}+ a_s^3\left(\mathcal{C}(\tau) - \mathcal{A}(\tau)\mathcal{B}(\tau) + \frac{1}{3}\mathcal{A}(\tau)^3 - G_{32}L^2 - G_{33}L^3 - G_{34}L^4\right)\,. \\
	\end{split}
\end{equation}
To ensure that $\Sigma_{NNLO+NNLL}(\tau_{max})=1$, where $\tau_{max}=$ is the maximum value $\tau$ can take with 5 final-state particles \cite{Cacciari:2025xoo}, we modify the logarithms as ~\cite{Jones:2003yv}
\begin{equation}
	L \to L' = \log\left(\left(\frac{1}{x_L y}\right) - \left(\frac{1}{x_Ly_\mathrm{max}}\right) + 1\right) \,
	\label{eq:modifiedLog}
\end{equation}
where the resummation scale $x_L$ is arbitrary.
\begin{figure}[t]
	\centering
	\includegraphics[width=0.24\textwidth]{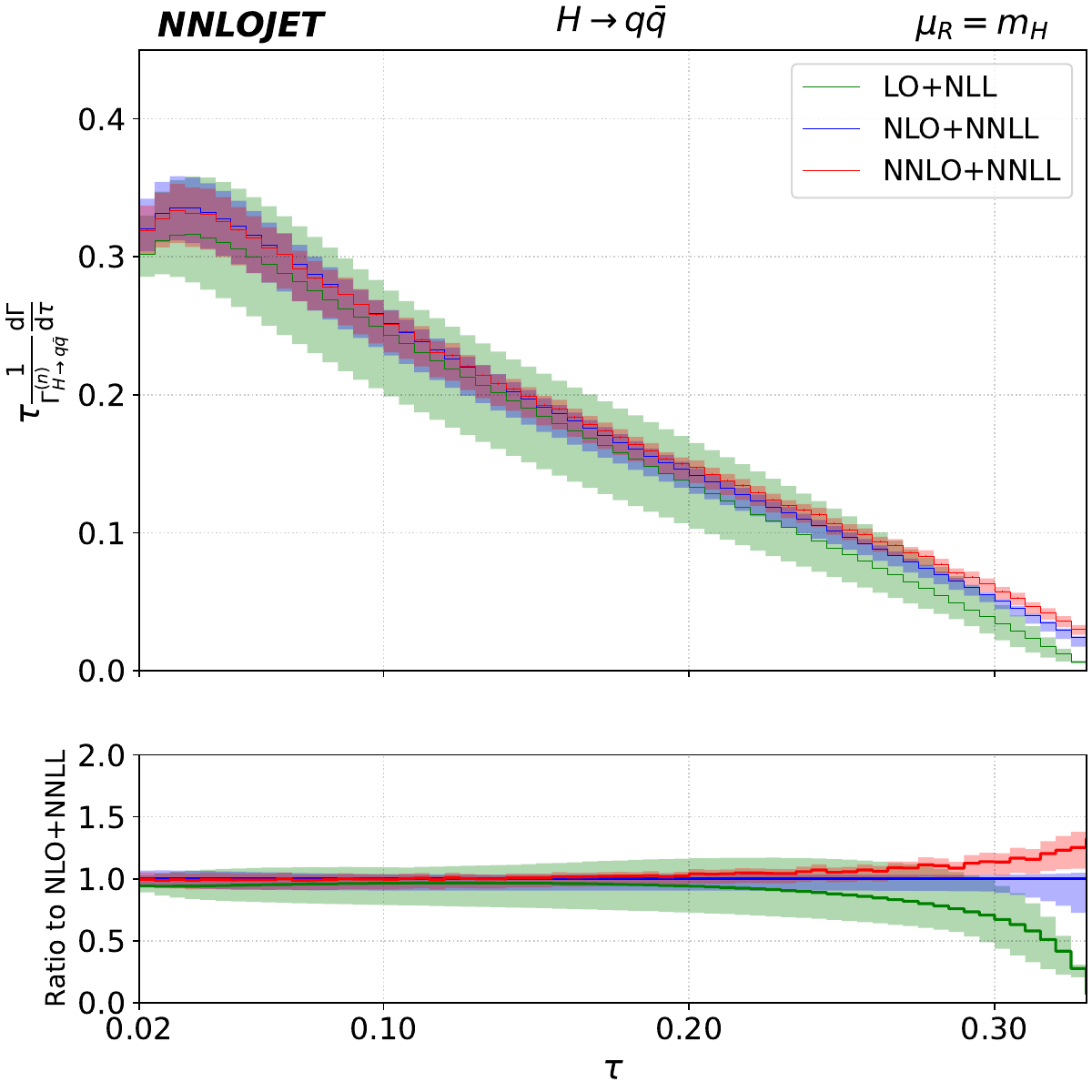}
	\includegraphics[width=0.24\textwidth]{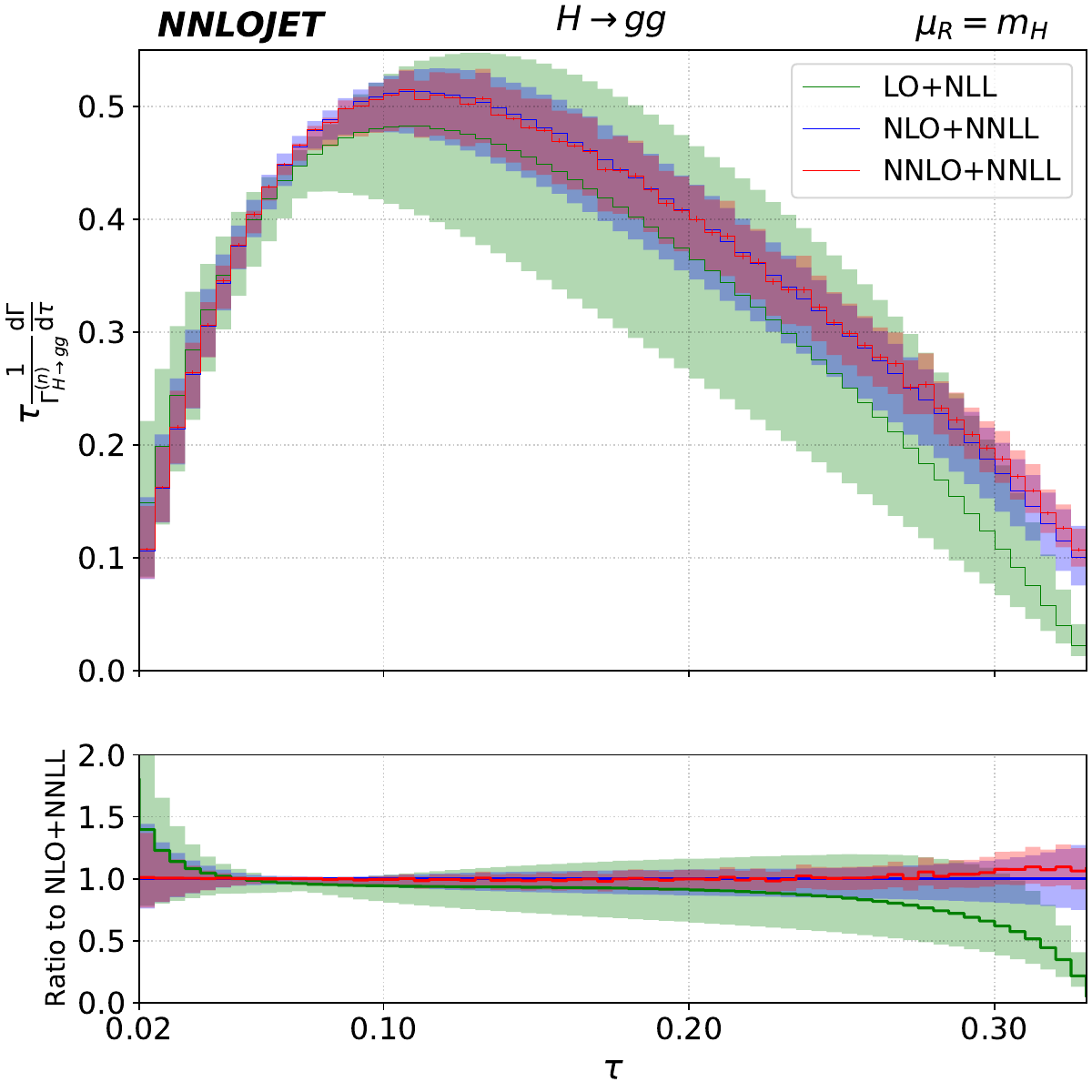}
	\includegraphics[width=0.24\textwidth]{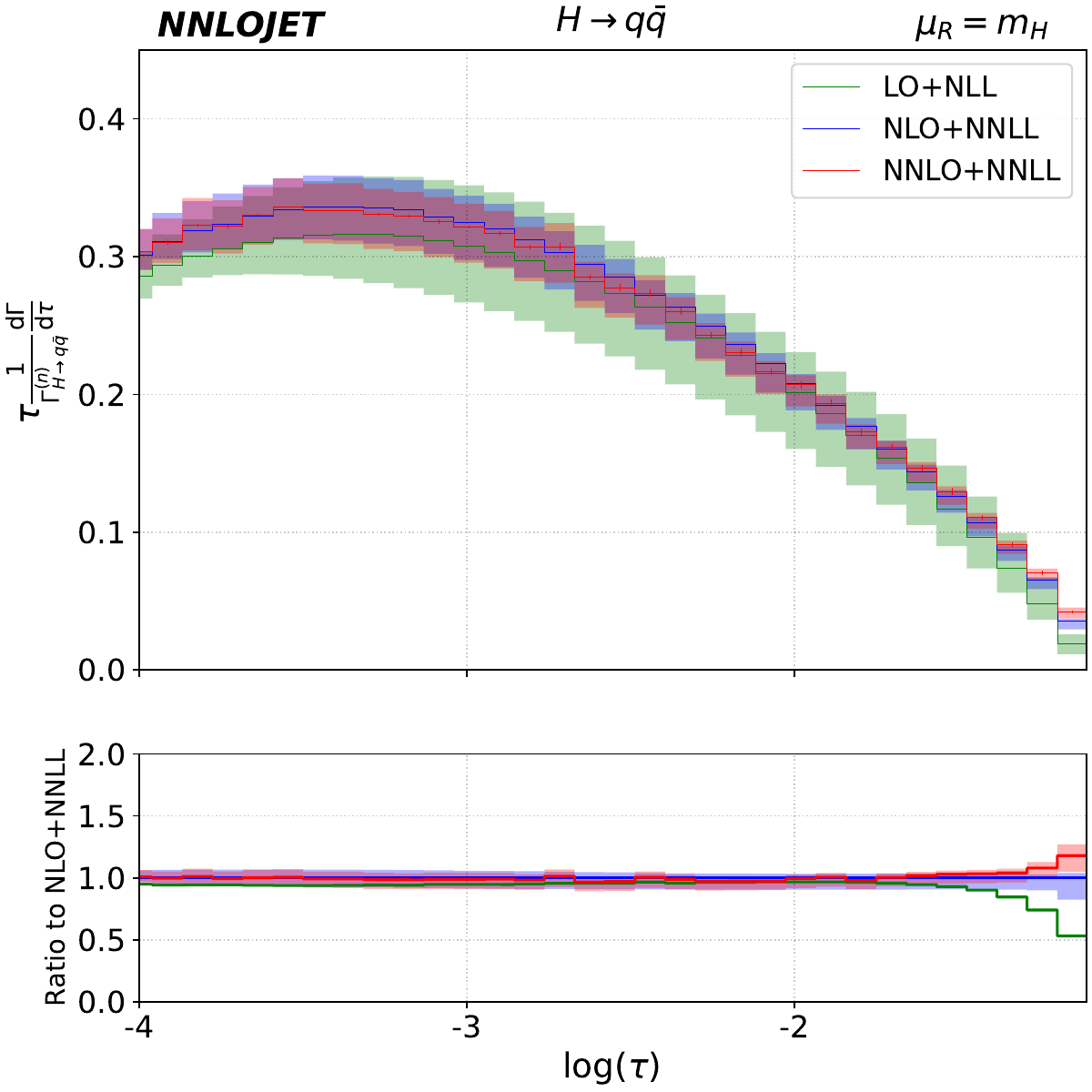}
	\includegraphics[width=0.24\textwidth]{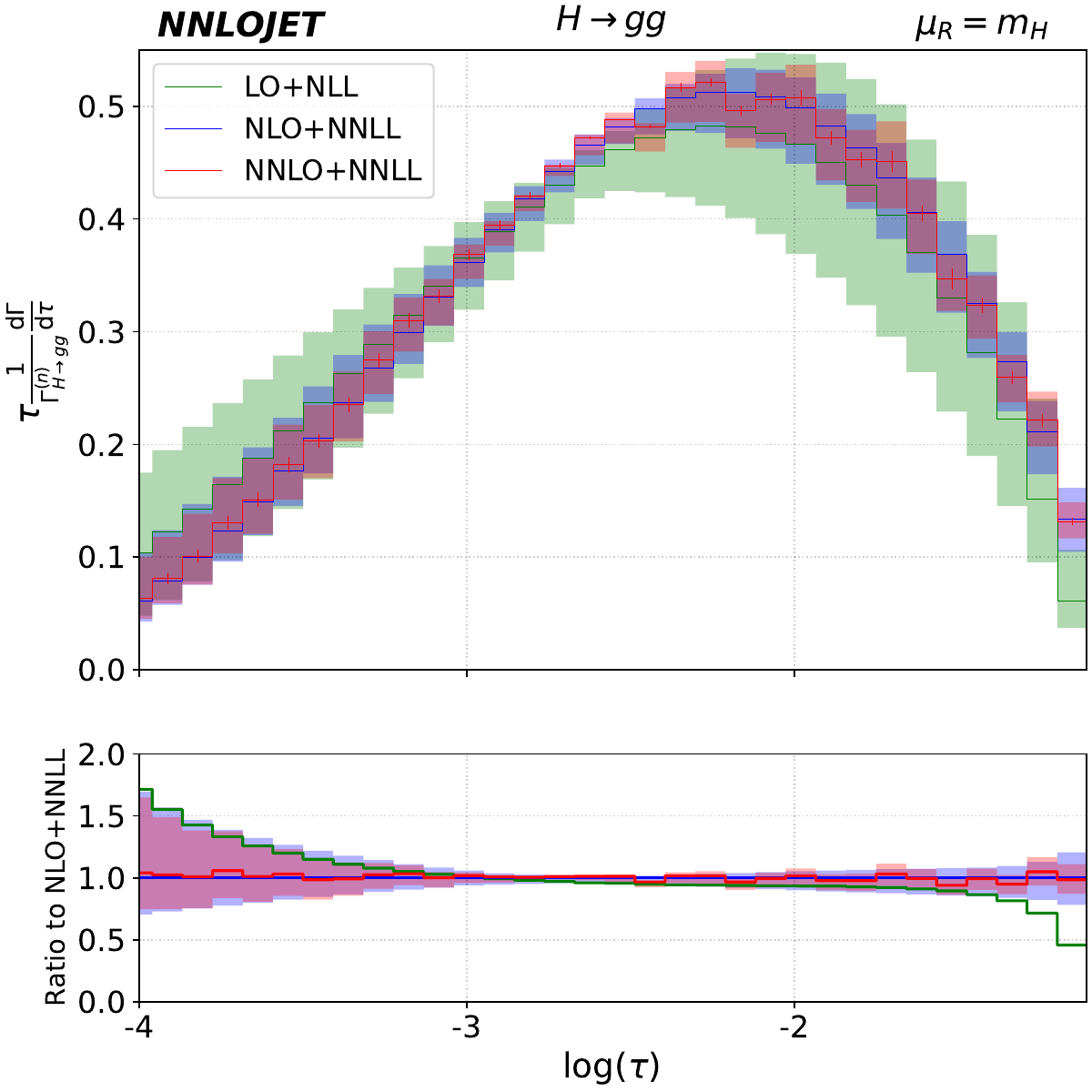}
	\caption{Thrust distribution results for fixed-order predictions matched to resummation in the logR scheme, for the $H\to q\bar{q}$ channel (left column) and the $H\to gg$ channel (right column), on a linear scale (top row) and on a log scale (bottom row). The curves represent the LO calculation matched to NLL resummation (green), NLO matched to NNLL (blue) and NNLO matched to NNLL (red). Vertical error bars indicate statistical Monte Carlo errors, while error bands show the combined scale variation}
	\label{fig:matched}
\end{figure}

\begin{figure}[t]
	\centering
	\includegraphics[width=0.26\textwidth]{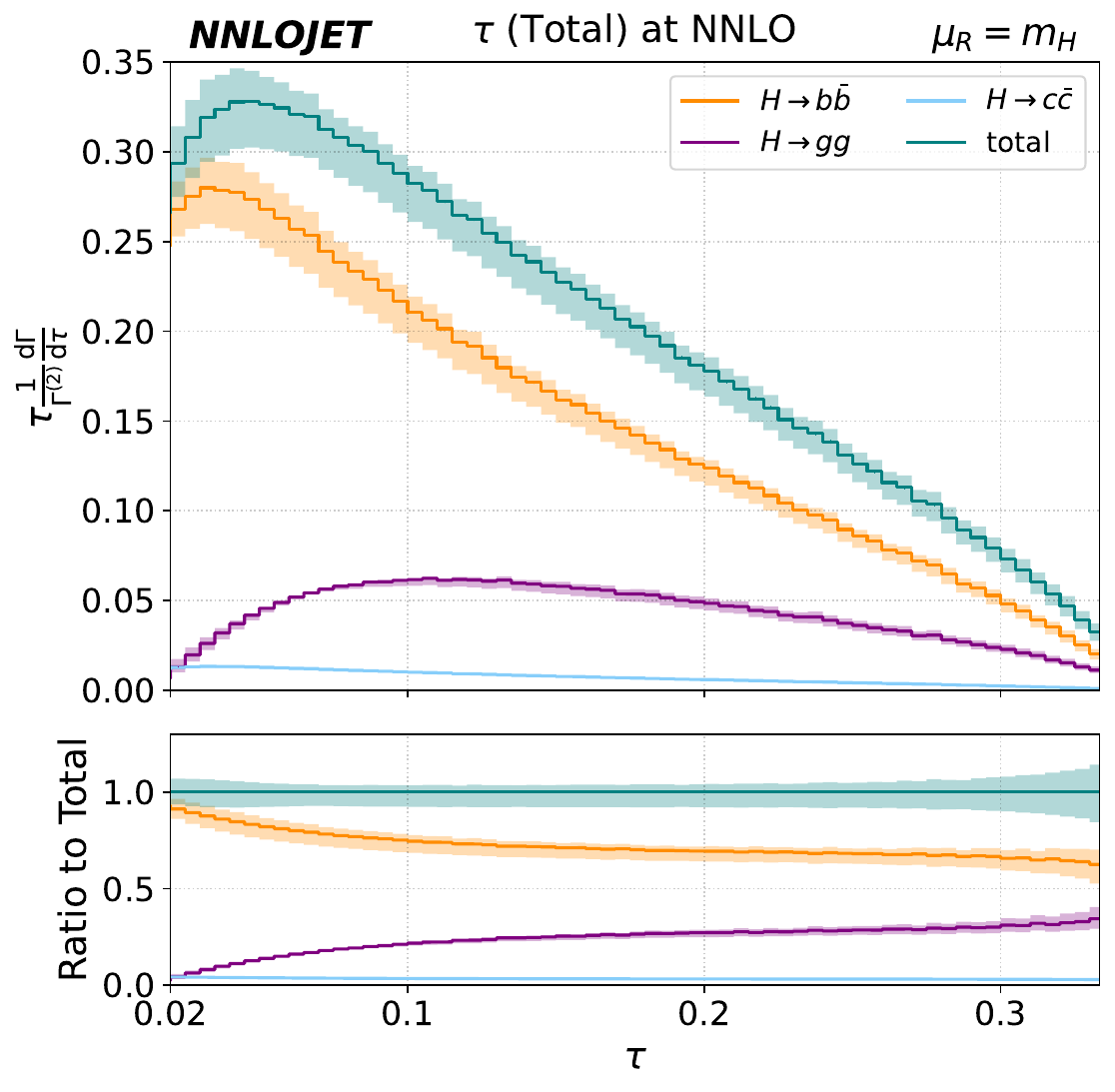}
	\includegraphics[width=0.26\textwidth]{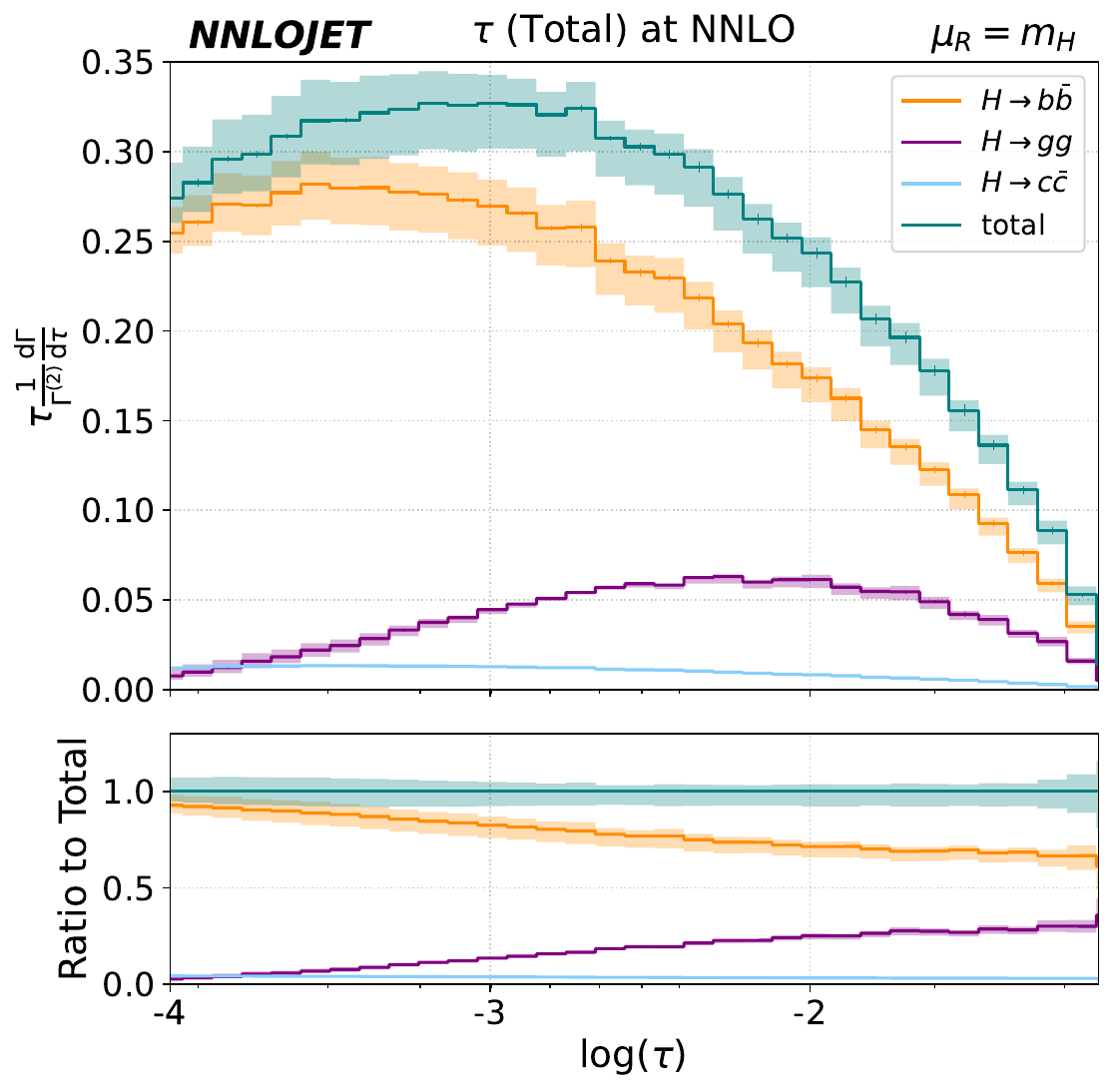}
	\caption{Results for the thrust distribution in the hadronic decay of a Higgs boson for a NNLO fixed-order calculation matched in the logR scheme to NNLL resummation. Curves for the total sum over all decay channels (teal), the $H\to b\bar{b}$ channel (orange), the $H\to gg$ channel (purple) and the $H\to c\bar{c}$ (light blue) are presented. The ratio to the total sum is shown in the lower frames. Vertical error bars indicate statistical Monte Carlo errors, while error bands show the combined scale variation.}
	\label{fig:matched_total}
\end{figure}

In Fig.\ref{fig:matched} we show matched predictions for the thrust observable in $H\to q\bar{q}$ and $H\to gg$ decays at LO+NLL, NLO+NNLL, and NNLO+NNLL, for both linear and logarithmic binning of $\tau$. To estimate theory uncertainties, we vary $\mu_R\in[m_h/2,2m_H]$ and $x_L\in[1/2,2]$, and from the overlapping uncertainty bands we see perturbative convergence has been restored. For both channels there is a value of $\tau$ at which the distribution peaks, and this peak is both higher and at a larger value of $\tau$ for $H\to gg$ than $H\to q\bar{q}$. The plots where $\tau$ is plotted linearly show that for large values of $\tau$, the fixed order dominates as the NLO+NNLL curves clearly differ from the NNLO+NNLL ones. On the other hand, the plots where $\tau$ is plotted logarithmically show that for small values of $\tau$ resummation dominates as the NLO+NNLL and NNLO+NNLL curves almost entirely overlap. In Fig.\ref{fig:matched_total} we show matched predictions at NNLO+NNLL for the $\tau$ distribution summed over all channels. We see that for large values of $\tau$ we inherit the features from the NNLO result in Fig.\ref{fig:tau_HJM}, while for $\tau<e^{-4.24}=0.0145$ the contribution from the $H\to gg$ channel vanishes, and the summed distribution consists only of $H\to b\bar{b}$ and $H\to c\bar{c}$ events. 

\section{Acknowledgements}
We would like to thank our collaborators Aude Gehrmann-De Ridder, Thomas Gehrmann, Nigel Glover, Matteo Marcoli and Christian Preuss.
\bibliographystyle{plain}
\bibliography{bibliography}
\end{document}